\documentclass[a4paper]{article}

\usepackage{latexsym}
\usepackage{amsmath}
\usepackage{amssymb}
\usepackage[dvips]{graphics}
\usepackage{cite}

\begin{document}

\def\manbo#1{}

\def\del{\partial}
\def\Lap{\bigtriangleup}
\def\goes{\rightarrow}
\def\inv{^{-1}}

\def\reff#1{(\ref{#1})}
\def\vb#1{{\partial \over \partial #1}} 
\def\vbrow#1{{\partial/\partial #1}} 
\def\Del#1#2{{\partial #1 \over \partial #2}}
\def\Dell#1#2{{\partial^2 #1 \over \partial {#2}^2}}
\def\Dif#1#2{{d #1 \over d #2}}
\def\Lie#1{ {\cal L}_{#1} }
\def\diag#1{{\rm diag}(#1)}
\def\abs#1{\left | #1 \right |}
\def\rcp#1{{1\over #1}}
\def\paren#1{\left( #1 \right)}
\def\brace#1{\left\{ #1 \right\}}
\def\bra#1{\left[ #1 \right]}
\def\angl#1{\left\langle #1 \right\rangle}

\def\S{\mathcal{S}}

\def\eh{e_2/H_0^2}
\def\ev{\frac{e_2}{H_0^2}}
\def\patT{1-H_0t_0}
\def\patLX{\frac{2-3H_0t_0}{\eh}}

 \def\manbo#1{}
 \newbox{\manbobox}
 \newenvironment{manboo}%
   {\begin{lrbox}{\manbobox}}%
   {\end{lrbox}}

\def\nospam{gravity.phys.nagoya-u.ac.jp}
\def\AT{@}

\title{Luminosity distance-redshift relation for the LTB solution
   near the center}

\author{Masayuki Tanimoto\thanks{E-mail: 
    \texttt{tanimoto\AT\nospam}} \  and 
  Yasusada Nambu\thanks{E-mail: 
    \texttt{nambu\AT\nospam}} \vspace{.5em}
\\
\textit{Department of Physics, Graduate School of Science,} \\
\textit{Nagoya University, Chikusa, Nagoya 464-8602, Japan}
}


\maketitle

{\abstract Motivated by the inverse problem for the
  Lema\^{i}tre-Tolman-Bondi dust solution, in which problem the
  luminosity distance function $D_L(z)$ is taken as an input to select
  a specific model, we compute the function $D_L(z)$ of the LTB
  solution up to the third order of $z$. To perform the otherwise
  cumbersome computation, we introduce a new convenient form of the
  LTB solution, in which the solution is explicit and unified. With
  this form of the LTB solution we obtain the luminosity distance
  function with the full generality. We in particular find that the
  function exactly coincides with that of a homogeneous and isotropic
  dust solution up to second order, if we demand that the solution be
  regular at the center.}

\section{Introduction}

Modern observations show that the Universe is presently in an
accelerating phase and dominated by dark energy with negative pressure
\cite{Riess:1998cb,Perlmutter:1998np,Riess:2004nr,
  Spergel:2003cb,Tegmark:2003ud}. It is remarkable that there seems no
obvious contradiction among the observations, but the true nature of
dark energy is a great mystery. On the other hand, it is also true
that this recognition is a result of the strict use of a
Friedmann-Lema\^{i}tre-Robertson-Walker (FLRW)
homogeneous and isotropic model. If inhomogeneities are properly taken
into account it might be possible to explain the observations without
introducing dark energy. Recently, this possibility has renewed
interest in inhomogeneous cosmological models, especially in the
so-called Lema\^{i}tre-Tolman-Bondi (LTB) solution \cite{L,T,B}, which
represents a spherically symmetric dust-filled universe. Because of
its simplicity this solution has been considered to be most useful to
evaluate the effect of inhomogeneities in the observables like the
luminosity distance-redshift relation.

Although this solution is simple, it has great flexibility. Various
models have been proposed that are consistent with the observations
\cite{Ce,Iguchi:2001sq,Alnes:2005rw,Bolejko:2005fp,Garfinkle:2006sb,
  Biswas:2006ub,Enqvist:2006cg}. But, what is the best configuration
that can explain the observations of, e.g., luminosity distance
redshift relation for type Ia supernovae and still be consistent with
other observations? To respond to this question one needs a
systematic approach.

Our primary focus is the ``inverse problem'' approach, in which one
takes a luminosity distance function $D_L(z)$ as an input to select a
specific LTB model.  Mustapha et.al. \cite{El98} argued that the
free functions in the LTB solution can be chosen to match any
given luminosity distance function $D_L(z)$ and source evolution
function $m(z)$. C\'el\'erier \cite{Ce} performed an expansion of
$D_L(z)$ for the LTB solution of parabolic type to fit the
$\Lambda$CDM model with $\Omega_\Lambda=0.7$ and $\Omega_M=0.3$, and
argued that the model can explain the SNIa observation at least for
$z\lesssim 1$. Vanderveld et.al. \cite{VFW} however showed that such a
fitting with an accelerating model is only possible at the cost of
occurrence of weak singularity at the center.

The main purpose of this paper is to find a Taylor expansion of
$D_L(z)$ for the LTB solution in the most general form, not restricted
to the parabolic type, with considerations for the condition to avoid
the weak singularity. Since the differential equation for the inverse
problem is in general singular at the origin $z=0$ it is imperative to
prepare a solution there using a different method. If we have the
expansion of $D_L(z)$ for the general LTB solution, we can easily find
this solution by comparing the given $D_L(z)$ and that of the LTB
solution order by order.

To perform the computation we present a new way of representing the
LTB solution. Although this is not our main purpose we would like to
stress that this new representation would be very useful in various
computations concerning the LTB solution. This solution is usually
expressed with parametric functions, as is the FLRW dust solution.
This parametric character of the solution often makes our analysis and
considerations complicated and non-transparent. Moreover those
functions change their forms depending on whether the solution is of
spherical type, parabolic type, or hyperbolic type. This
split-into-cases character is also a drawback of the conventional
expression of the solution.  The new expression of the solution
dissolves all these unwanted characters. It will play an essential
role to achieve the complicated computations needed to compute
$D_L(z)$ in the fully general setting.

The structure of this paper is as follows. In section \ref{sec:2} we
introduce the new unified form of the LTB solution. In section
\ref{sec:3} we perform the expansion of $D_L(z)$ using the unified
expression. In section \ref{sec:4} we study how different the function
$D_L(z)$ for the LTB solution is from that of the FLRW dust
solution. Section \ref{sec:conc} is devoted to conclusion. Appendix
\ref{app1} presents the result without the regularity condition at the
center.

\section{The LTB solution in a unified new form}
\label{sec:2}

Let us first gather the conventional forms of the solution.
The LTB metric is written in the form
\begin{equation}
  \label{eq:metric}
  ds^2=-dt^2+\frac{{R'(t,r)}^2}{1+2E(r)}dr^2+
  R(t,r)^2(d\theta^2+\sin^2\theta d\phi^2),
\end{equation}
where primes denote derivatives with respect to the radial coordinate
$r$, and $E(r)$ is a free function called the ``energy function''. The
conventional way of expressing the areal radius function
$R(t,r)$ depends on the sign of $E(r)$ and is parametric:
For $E(r)>0$, we have
\begin{align}
  \label{eq:E>0}
  R(t,r) &= \frac{M(r)}{2E(r)}(\cosh\eta-1), \\
  \label{eq:E>0t}
  t-t_B(r) &= \frac{M(r)}{(2E(r))^{3/2}}(\sinh\eta-\eta),
\end{align}
where $M(r)$ and $t_B(r)$ are free functions called, respectively, the
``mass function'' and the ``big bang function''. For $E(r)<0$, we have
\begin{align}
  R(t,r) &= \frac{M(r)}{-2E(r)}(1-\cos\eta), \\
  t-t_B(r) &= \frac{M(r)}{(-2E(r))^{3/2}}(\eta-\sin\eta).
\end{align}
Finally, for $E(r)=0$, we have
\begin{equation}
  \label{eq:E=0}
  R(t,r)=\paren{\frac92}^{1/3}M(r)^{1/3}(t-t_B(r))^{2/3}.
\end{equation}
(In this case the function $R(t,r)$ is explicit in terms of $t$ and $r$.)

Now, observe that in the $E>0$ case, Eq.\reff{eq:E>0t} shows that
$\eta$ can be regarded as a function of
$2E((t-t_B)/M)^{2/3}$. Therefore from Eq.\reff{eq:E>0} the function
$R$ can be written in the form
\begin{equation}
  R(t,r)= \frac{M}{2E}X\paren{2E\paren{\frac{t-t_B}{M}}^{2/3}},
\end{equation}
using a certain function $X(x)$. To make the $E\goes0$ limit
apparently regular, we then factor out $x$ from the function
$X(x)$ and make $X(x)=6^{1/3} x \S(-6^{-2/3}x)$;
\begin{equation}
  \label{eq:RS}
  R(t,r)= (6 M(r))^{1/3}(t-t_B(r))^{2/3}
  \S\paren{-2 E(r)\paren{\frac{t-t_B(r)}{6M(r)}}^{2/3}}.
\end{equation}
(The numerical factors inserted are our convention.)

The same observation is also applicable to both $E<0$ and $E=0$ cases,
and as a result, we find that the form \reff{eq:RS} may provide a
desirable form of $R(t,r)$, which is explicit in terms of $t$ and $r$
and requires no separate considerations depending on the sign of
$E(r)$. There is, however, still a remaining task, which is to confirm
that the function $\S(x)$ is smooth at $x=0$. Otherwise, this form
would be superficial.

\begin{figure}[hbtp]
  \begin{center}
\scalebox{.8}{\includegraphics{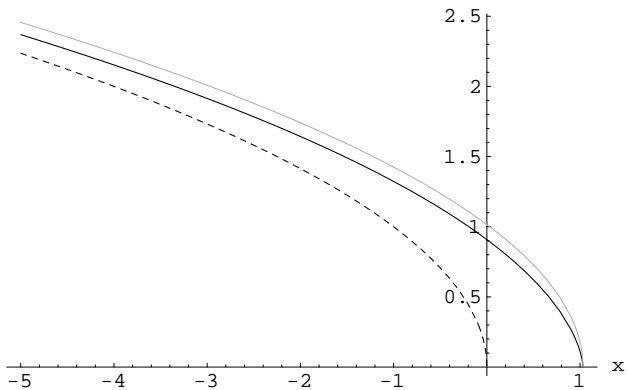}}
  \end{center}
  \caption{Plots of the function $\S(x)$ (solid line), the
    function $\mathcal{\tilde V}_0(x)=\sqrt{-x}$ (dashed line), and
    the function $\S_C(x)=\sqrt{x_C-x}$ (gray line). Although the
    parametric expression of $\S(x)$ takes different forms for $x>0$,
    $x=0$, and $x<0$, this function is smooth at the dividing point
    $x=0$.}
  \label{fig:1}
\end{figure}

To this, note that the function $\S(x)$ can be expressed in parametric
forms.  For $x<0$ (corresponding to $E>0$), we have
\begin{equation}
  \label{eq:Sx<0}
  \S(x) = \rcp{6^{1/3}}
  \frac{\cosh\sqrt{-\zeta}-1}{(\sinh\sqrt{-\zeta}-\sqrt{-\zeta})^{2/3}},\quad
  x = \frac{-1}{6^{2/3}}(\sinh\sqrt{-\zeta}-\sqrt{-\zeta})^{2/3}.
\end{equation}
For $x>0$ ($E<0$), we have
\begin{equation}
  \label{eq:Sx>0}
  \S(x) = \rcp{6^{1/3}}
  \frac{1-\cos\sqrt{\zeta}}{(\sqrt{\zeta}-\sin\sqrt{\zeta})^{2/3}},\quad
  x = \rcp{6^{2/3}}(\sqrt{\zeta}-\sin\sqrt{\zeta})^{2/3}.
\end{equation}
For $x=0$, we have
\begin{equation}
  \label{eq:S0}
  \S(0)=\paren{\frac34}^{1/3}.
\end{equation}
The parameter $\zeta$ takes positive and negative values in accordance
with the sign of $x$. (The explicit correspondence to $\eta$ is given
by $\zeta=\eta^2$ for $E<0$, and $\zeta=-\eta^2$ for $E>0$.)
We then find that $\S$ and $x$ are expanded in powers of $\zeta$ in
the same form for all signs of $x$, since by direct computation we can
immediately confirm the following common expression:
\begin{equation}
  \label{eq:Sxzetapro}
  \S(x) = \rcp{6^{1/3}}
  \frac{\frac12-\frac1{4!}\zeta+\cdots}{
    \paren{\frac1{3!}-\frac{\zeta}{5!}+\cdots}^{2/3}},\quad
  x = \frac{\zeta}{6^{2/3}}\paren{
    \frac1{3!}-\frac{\zeta}{5!}+\cdots}^{2/3}.
\end{equation}
Proceeding the series expansions,
\begin{equation}
  \label{eq:Sxzeta}
\begin{split}
  \S(x) & =
  \paren{\frac34}^{1/3}\paren{1-\frac{\zeta}{20}+\frac{\zeta^2}{1680}
  +O(\zeta^3)}, \\
  x &= 6^{-4/3}\paren{\zeta-\frac{\zeta^2}{30}+\frac{13\zeta^3}{25200}
    +O(\zeta^4)}.
\end{split}
\end{equation}
Since both $\S$ and $x$ are smooth (in fact, analytic) functions of
$\zeta$, in particular in the neighborhood of $\zeta=0$, this shows
that $\S(x)$ is smooth at $x=0$, as desired. (The smoothness of
$\S(x)$ at $x\neq0$ is apparent.)  Figure \ref{fig:1} shows a plot of
the function $\S(x)$. $\S(x)$ is a non-negative, monotonic function,
defined for $x\leq x_C\equiv(\pi/3)^{2/3}$.

The expression \reff{eq:RS} therefore does provide a desired form of
the LTB solution, together with the conventional metric form
\reff{eq:metric}. This expression will turn out to be extremely useful
in performing most of the computations concerning the LTB solution,
including the expansion of $D_L(z)$ in the next section. In the rest
of this section we briefly discuss some useful properties concerning
the function $\S(x)$.

First, remember that $R(t,r)$ satisfies the following generalized
`Friedmann' equation:
\begin{equation}
  \label{eq:RF}
  {\dot R}^2= \frac{2M}{R}+2E,
\end{equation}
where dots denote derivatives with respect to the proper time
$t$. Substituting Eq.\reff{eq:RS} we immediately have the first order
ordinary differential equation (ODE) for $\S(x)$:
\begin{equation}
  \label{eq:SODE1}
  \frac43(\S(x)+x \S'(x))^2+3x-\rcp{\S(x)}=0,
\end{equation}
where $\S'\equiv d\S(x)/dx$. (We understand that primes attached to
$\S$ always stand for derivatives with respect to its single argument,
\textit{not} to the radial coordinate $r$). 


This ODE has a peculiar feature observed at $x=0$; at this point the
term $x\S'(x)$ vanishes and therefore the equation degenerates into an
algebraic equation that constrains $\S(0)$, provided that
$|\S'(0)|<\infty$. As a result we obtain $S(0)=(3/4)^{1/3}$ in
accordance with Eq.\reff{eq:S0}. The finiteness condition of $S'(x)$
at $x=0$ implies that the line $y=S(x)$ in the $x$-$y$ plane should
intersect the $x=0$ axis transversely. The above feature therefore
tells us that the function $S(x)$ is \textit{the unique `transversal'
  solution of the ODE \reff{eq:SODE1}}. It may be useful to use this
characterization to define $S(x)$, instead of using the explicit
parametric expressions \reff{eq:Sx<0}, \reff{eq:Sx>0}, and
\reff{eq:S0}.

\begin{figure}[hbtp]
  \begin{center}
\scalebox{.4}{\includegraphics{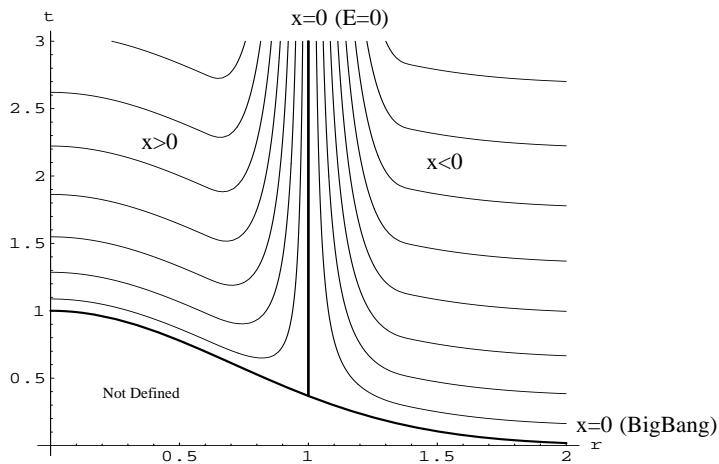}}
  \end{center}
  \caption{Typical contours of $x(t,r)$ when $E(r)$ crosses zero. 
    The decreasing bold curve corresponds to the Big Bang singularity
    ($t=t_B(r)$), while the vertical bold line to $E(r)=0$
    (this is \textit{not} a singularity). Both bold lines correspond to
    $x=0$. Contours are drawn for $t\geq t_B(r)$. [This example is
    intended to be helpful for a general discussion of the properties
    of $\S(x)$, and not to imply a setting of the next section. The
    specific choice for this example is
    $E(r)=-r^2(1/2-\theta_{0.4}(r-1))$, $M(r)=r^3$, and
    $t_B(r)=e^{-r^2}$. The function $\theta_\sigma(r)$ is a step
    function with the transition interval $[-\sigma,\sigma]$ with
    values $\theta_\sigma(r)=1 (r\geq\sigma)$, and $0
    (r\leq-\sigma)$.]}
  \label{fig:2}
\end{figure}

It may be worth commenting that we should not think of the ODE
\reff{eq:SODE1} as a usual `evolution equation,' because of the
following two reasons.  First, as discussed previously, this ODE has
no freedom to choose an initial value. Note that the solution of the
generalized Friedmann equation \reff{eq:RF} does possess freedom to
choose an initial data, which is the function $t_B(r)$. This function
is however already incorporated in the form \reff{eq:RS}. The
resulting equation therefore cannnot contain further freedom to choose
a solution. Second, the gradient of the variable $x(t,r)$ for the ODE
is in general not timelike. In other words, the contours of $x(t,r)$
are not always spacelike. This is most noticeable when $E(r)$ crosses
zero at some spatial points. Figure \ref{fig:2} shows an example where
$E(r)$ crosses zero at one point ($r=1$).

Note that the form \reff{eq:RS} already contains the parabolic
solution as one factor, with the remaining factor being $S(x)$. We can
think of $|x|$ as a parameter that measures the deviation from the
parabolic evolution. It is obvious that $x=0$ corresponds to both big
bang time $t=t_B(r)$ and comoving points $E(r)=0$. In the above
example, these two regions are shown as the bold lines forming an
inverted `T'-shape (Fig.\ref{fig:2}). Let us consider a neighborhood
of this `T'-shaped region such that for a small positive $\epsilon$,
$|x|<\epsilon$. Then we can say that the time evolution of the points
contained in this spacetime region is close to that of the parabolic
solution.  Actually, for small $t-t_B(r)$ this is a well-known
fact. (Both elliptic and hyperbolic solutions asymptotically approach
the parabolic type evolution $(t-t_B(r))^{2/3}$ near the big bang
time.)  It is also apparent that if $|E(r)|$ is small enough the
evolution of those points is close to that of the parabolic solution
(for a finite time-interval). The variable $x$ is therefore a
deviation parameter, rather than a evolution parameter.


Now, let us return to the discussion of useful properties of
$\S(x)$. Remember that Eq.\reff{eq:RF} is an integral of the following
second order partial differential equation:
\begin{equation}
  \label{eq:RODE2}
  2R\ddot{R}+\dot{R}^2-2E=0.
\end{equation}
From this we have the second order ODE for $\S(x)$:
\begin{equation}
  \label{eq:SODE2}
  x(2\S(x)\S''(x)+{\S'}^2(x))+5\S(x)\S'(x)+\frac94=0.
\end{equation}
For $x\neq0$, this equation is useful in case we want to eliminate
higher derivatives of $\S(x)$ than first order.

When $x=0$, Eq.\reff{eq:SODE2} itself does not determine $\S''(0)$,
due to the multiplied factor $x$. However, we can regard this equation
as an equation that determines $\S'(0)$ from $\S(0)$, and find
$\S'(0)=-9/(20\S(0))$. Similarly, taking derivatives of the equation
we can successively determine the values of arbitrarily higher
derivatives of $\S(x)$ at $x=0$, in terms of $\S(0)$. One of the
useful applications of this property is to obtain a series expansion
of $\S(x)$ about $x=0$. Since $\S(0)$ is given in Eq.\reff{eq:S0} (or
determined from Eq.\reff{eq:SODE1}), we have
\begin{equation}
  \label{eq:Span}
  \S(x)=\paren{\frac34}^{1/3}\paren{
    1-\frac{3}{5}\paren{\frac{3}{4}}^{1/3}x
    -\frac{27}{350}\paren{\frac92}^{1/3}x^2 + O(x^3)
  }.
\end{equation}
As discussed previously, this expansion can provide a `parabolic
approximation' of the solution.

We may be interested in the significance of the general solutions of
the ODE \reff{eq:SODE2}. To avoid confusions let us top tilde, like
$\tilde\S(x)$, to denote a solution of the ODE which does not
necessarily coincide with $\S(x)$. First, the above observation
immediately tells us that a solution of the ODE which intersects the
$x=0$ axis transversely is specified only by one parameter
$s\equiv\tilde\S(0)$, since ${\tilde\S}'(x)$ is not free at $x=0$. We
can easily confirm that all those transversal solutions are generated
from $\S(x)$ by rescaling: \begin{equation}
  \tilde\S_\alpha(x)=\alpha\S\paren{\frac x{\alpha^2}}, \end{equation}
where $\alpha=s/\S(0)=(4/3)^{1/3}s$. This rescaled function is also a
solution of the first order ODE \reff{eq:SODE1} with the modification
$1/\S \goes 1/(\alpha^3\S)$. These rescaling of the function and
modification of the equation just correspond to the rescaling
$M(r)\goes \alpha^3 M(r)$, and therefore although the rescaled
function $\tilde\S_\alpha(x)$ does generate a dust solution, it is
equivalent to the one by the original $\S(x)$. Thus, we may not be
interested in the general transversal solution $\tilde\S_\alpha(x)$.

Non-transversal solutions of the ODE \reff{eq:SODE2} in general have
two-parameters. An interesting one-parameter special solution is
\begin{equation}
  \mathcal{\tilde V}_\beta(x)= \sqrt{-x}-\frac{\beta}{x},
\end{equation}
where $\beta$ is an arbitrary constant parameter.
$\mathcal{\tilde V}_\beta(x)$ approaches $\S(x)$ as $x\goes -\infty$:
\begin{equation}
  \mathcal{\tilde V}_\beta(x)\goes \S(x) \quad (x\goes -\infty).
\end{equation}
In particular, $\mathcal{\tilde V}_0(x)=\sqrt{-x}$ approaches $\S(x)$
from below (Fig.\ref{fig:1}), and is often useful for various
estimates of $\S(x)$ for large $-x$.  We remark that the function
$\mathcal{\tilde V}_\beta(x)$ satisfies the first order ODE
\reff{eq:SODE1} if the $1/\S(x)$ term in the equation is
neglected. This in effect corresponds to taking the limit $M(r)\goes
0$, and therefore the metric generated by this function (through
Eq.\reff{eq:RS} with $\S(x)$ replaced by $\mathcal{\tilde
  V}_\beta(x)$) represents a vacuum solution. In fact, by a direct
computation of the Riemann tensor we find that it all represents
the Minkowski solution.

Finally, it is worth mentioning that,  slightly modifying
$\mathcal{\tilde V}_0(x)$, the function
\begin{equation}
  \S_C(x)\equiv \sqrt{x_C-x},
\end{equation}
which is no longer a solution of any of the ODEs, provides a good
approximation\footnote{It is interesting to note that in the special
  case of the FLRW dust solution, the approximated solution with
  $\S_C(x)$ corresponds to the ``renormalized solution''
  \cite{Nambu:1999aw} of a solution obtained in the long wavelength
  approximation of Einstein's equation.} of $\S(x)$ for all domain of
$x\leq x_C$. The function $\S_C(x)$ asymptotically approaches $\S(x)$
from above as $x\goes -\infty$ (Fig.\ref{fig:1}). An elementary
estimate actually confirms the inequality
\begin{equation}
  \S(x)\leq \S_C(x),
\end{equation}
where the equality holds for $x=x_C$. Also, another elementary estimate
establishes
\begin{equation}
  \label{eq:S'ineq}
  x\S'(x)<\frac{\S(x)}{2}
\end{equation}
for the first derivative.

\section{Expansion of $D_L(z)$}
\label{sec:3}

Motivated by the inverse problem, we in this section perform the
expansion of the function $D_L(z)$, the luminosity distance as a
function of the redshift, for the LTB solution.

The luminosity distance for the LTB solution is simply given by
\cite{Ellis:1971pro,PM84}
\begin{equation}
  \label{eq:DLR}
  D_L=(1+z)^2 R,
\end{equation}
where the areal radius function $R=R(t(z),r(z))$ must be
evaluated along the light ray emitted from the light source and caught
by the central observer at $(t,r)=(t_0,0)$. With this formula we can
expand $D_L(z)$ if we expand $R$.

Before proceeding the expansion, we note that the LTB solution has
the gauge freedom of choosing the radial coordinate $r$. In
fact, it is easy to see that coordinate transformation $r\goes \tilde
r =\tilde r(r)$ retains the characteristic form of the solution if we
redefine the free functions $E(r), t_B(r)$, and $M(r)$ suitably. The
significance of this freedom is that it enables us to fix the function
$M(r)$ for many cases.

One of the popular gauge choices is (e.g.,
\cite{Ce,Biswas:2006ub,VFW}) to take  $M(r)=M_0 r^3$ with $M_0$ being
a positive constant. We call this choice the \textit{FLRW gauge},
since the standard form of the FLRW dust solution is realized with
this choice with the other functions being $E(r)=-(k/2)r^2$, and
$t_B(r)=0$. ($k$ is the curvature constant.)

Another excellent choice is \cite{El97,El98} to choose the radial
coordinate $r$ so that the light rays coming into the observer are
simply expressed by
\begin{equation}
  \label{eq:rays}
  t=t_0-r.
\end{equation}
This is equivalent to the condition that $dr/dt=-1$ must hold along
those light rays, and therefore from the metric \reff{eq:metric} it is
equivalent to imposing
\begin{equation}
  \label{eq:Elgg}
  \frac{R'(t_0-r,r)}{\sqrt{1+2E(r)}}=1,
\end{equation}
where $R'(t_0-r,r)\equiv (\partial R(t,r)/\partial
r)|_{t=t_0-r}$. This gauge choice determines the function $M(r)$ only
simultaneously with $E(r)$ and $t_B(r)$. We call this choice of $r$ or
the resulting choice of $M(r)$ the \textit{light cone gauge}.

We in the following adopt the light cone gauge, which, together with
the formula \reff{eq:RS}, makes our procedure of expansion
systematic and straightforward. In fact, with the condition
\reff{eq:rays} we can firstly expand $R$ in powers of $r$ in a
straightforward manner, putting $R=R(t_0-r,r)$ in the formula
\reff{eq:RS}. To convert the result into expansion in terms of $z$, we
need to find $r(z)$ in powers of $z$. This is possible from the
formula\cite{Ce}
\begin{equation}
  \label{eq:drdz}
  \frac{dr}{dz}=\frac{\sqrt{1+2E(r)}}{(1+z){\dot R}'(t_0-r,r)}.
\end{equation}

This would complete our expansion, but we still need to consider an
extra condition, which is the regularity at $r=0$. As pointed out in
\cite{VFW,Mustapha:2000bf}, if the first derivative with
respect to $r$ of the matter density $\rho(t,r)$ does not vanish at
$r=0$, the spacetime has a weak singularity there. We wish to
eliminate this undesirable feature; therefore impose
\begin{equation}
  \label{eq:reg0}
  \rho'(t,0)=0.
\end{equation}
The expansion with this condition imposed will give the final form of
our result.

We are now in a position to proceed the expansion, following the
procedure outlined above. We wish to expand $D_L(z)$ up to the third
order, which leads us to expand the free functions as follows:
\begin{equation}
  \label{eq:fexp}
  \begin{split}
    E(r) &=
    \frac{e_2}{2}r^2+\frac{e_3}{3!}r^3+\frac{e_4}{4!}r^4+O(r^5), \\
    t_B(r) &=
    {b_1}r+\frac{b_2}{2}r^2+O(r^3), \\
    M(r) &=
    \frac{m_3}{3!}r^3+\frac{m_4}{4!}r^4+\frac{m_5}{5!}r^5+O(r^6),
  \end{split}
\end{equation}
where the differential coefficients $e_i\equiv E^{(i)}(0)$, 
$b_i\equiv t_B^{(i)}(0)$, and $m_i\equiv M^{(i)}(0)$ are
constants, with $m_3>0$. Our task is then to express the
expansion of $D_L(z)$ in terms of these coefficients.

It turns out to be convenient to, prior to the general computations,
finish the computation for the first order of $D_L(z)$ to determine
the Hubble constant $H_0$ in comparison with the definition
$D_L(z)=z/H_0 +O(z^2)$. This is not very much elaborating, and we find
\begin{equation}
  \label{eq:H0}
  H_0=\frac{2}{3t_0}\paren{1+x_0 \frac{\S'(x_0)}{\S(x_0)}},
  \quad x_0\equiv -e_2\paren{\frac{t_0}{m_3}}^{2/3}.
\end{equation}
(As remarked, prime attached to $\S(x)$ stands for the derivative with
respect to $x$, not with respect to $r$; $\S'(x_0)\equiv
(d\S(x)/dx)|_{x=x_0}$.)

The gauge condition \reff{eq:Elgg} is equivalent to the vanishing of
the function
\begin{equation}
  \label{eq:G}
  G(r)\equiv \sqrt{1+2 E(r)}-R'(t_0-r,r).
\end{equation}
We expand this function up to second order and demand that the
coefficients be equal to zero. First of all, from the 0th coefficient
we have
\begin{equation}
  \label{eq:Srl}
  \S(x_0)=t_0^{-2/3}m_3^{-1/3},
\end{equation}
and together with Eq.\reff{eq:H0}, we also have
\begin{equation}
  \label{eq:Sdrl}
  \S'(x_0)=\frac{(2-3H_0t_0)m_3^{1/3}}{2e_2t_0^{4/3}}.
\end{equation}
In the following we will use Eqs.\reff{eq:Srl} and \reff{eq:Sdrl} to
eliminate $\S(x_0)$ and $\S'(x_0)$ from various equations.  We remark
that as we will see, Eq.\reff{eq:Sdrl} is valid not only for the case
$e_2\neq0$ but also for the $e_2=0$ case, if we understand that an
appropriate limit is taken. Note also that Eqs.\reff{eq:Srl} and
\reff{eq:Sdrl} are not independent, since so are not $\S(x)$ and
$\S'(x)$. Substituting the equations into Eq.\reff{eq:SODE1} we obtain
the relation
\begin{equation}
  \label{eq:e2m3}
  \frac{e_2}{H_0^2}=1-\frac{m_3}{3H_0^2}.
\end{equation}
Since $m_3>0$, this also means
\begin{equation}
  \label{eq:e2c}
  \frac{e_2}{H_0^2}<1.
\end{equation}

To take the $e_2\goes 0$ limit in various equations that will appear
in the rest of our computation, it is useful to have an expression
of $H_0t_0$ in powers of $e_2/H_0^2$. First, from Eqs.\reff{eq:H0}
and \reff{eq:Span} we have
\begin{equation}
  \label{eq:H0t0pan}
  H_0t_0=\frac23\paren{1-\frac35\paren{\frac34}^{1/3}x_0
    -\frac{117}{350}\paren{\frac92}^{1/3}x_0^2+O(x_0^3)}.
\end{equation}
On the other hand, from Eq.\reff{eq:e2m3} (and the definition of $x_0$
\reff{eq:H0}) we have
\begin{equation}
  x_0=-\frac{e_2}{H_0^2}\paren{\frac{H_0t_0/3}{1-e_2/H_0^2}}^{2/3}.
\end{equation}
Substituting this into Eq.\reff{eq:H0t0pan}, we obtain an equation
for $H_0t_0$. We can solve this and find
\begin{equation}
  H_0t_0=\frac23\paren{1+\frac15\frac{e_2}{H_0^2}
    +\frac{3}{35}\paren{\frac{e_2}{H_0^2}}^2 +O\paren{(e_2/H_0^2)^3}}.
\end{equation}
This is the expression we wanted. In particular, this implies as
$e_2\goes0$,
\begin{align}
  \label{eq:e20limiteqn}
  \frac{2-3H_0t_0}{e_2/H_0^2} &\goes -\frac{2}{5}, \\
  \label{eq:e20limiteqn2}
  \paren{\frac{2-3H_0t_0}{e_2/H_0^2}+\frac{2}{5}}\rcp{e_2/H_0^2}
  & \goes -\frac{6}{35}.
\end{align}
With the first of these we can confirm that Eq.\reff{eq:Sdrl} provides
the right value for the limit $e_2\goes0$, and therefore
Eq.\reff{eq:Sdrl} is, as mentioned, valid also for $e_2=0$.

Returning to the gauge condition \reff{eq:G}, from its first order
coefficient we have
\begin{equation}
  \label{eq:m4rl}
  m_4 =
  \frac{4m_3}{t_0}-\frac{4e_2m_3^{1/3}\S'(x_0)}{t_0^{1/3}\S(x_0)} \\
  = 6 H_0 m_3.
\end{equation}
Similarly, from the second order we have
\begin{equation}
  \label{eq:m5rl}
  m_5=\frac{m_3}{2}\paren{15H_0^2+5\frac{e_4}{e_2}
  +\frac{5}{3(1-H_0t_0)}\paren{13 H_0^2+12 H_0b_2+2e_2-\frac{e_4}{e_2}}}.
\end{equation}
(The second derivative $\S''(x_0)$ appears in the coefficient, which
we eliminate using Eq.\reff{eq:SODE2}.) In particular, for $e_2=0$ we
have
\begin{equation}
  \label{eq:m5rl0}
  m_5= m_3 \paren{10 H_0(4H_0+3b_2)-\frac{e_4}{H_0^2}}.
\end{equation}
In the following we will eliminate $m_4$ and $m_5$ with
Eqs.\reff{eq:m4rl} and \reff{eq:m5rl}.

To find the regularity condition at $r=0$ we expand
\begin{equation}
  4\pi\rho(t,r)=\frac{M'(r)}{R^2(t,r)R'(t,r)}
\end{equation}
in powers of $r$ up to first order with $t$ fixed:
\begin{equation}
  \label{eq:rho1st}
  \begin{split}
  4\pi\rho(t,r)= &
  \frac1{\S^3(x(t))}\bigg[ \frac{1}{2t^2} \\
  &  +\paren{\frac{4b_1}{3t^3}
    -\frac{4b_1e_2}{3m_3^{2/3}t^{7/3}}\frac{\S'(x(t))}{\S(x(t))}
    +\frac{(2e_3m_3-e_2m_4)}{3m_3^{5/3}t^{4/3}}\frac{\S'(x(t))}{\S(x(t))}}r 
  \\
  & +O(r^2) 
  \bigg],
  \end{split}
\end{equation}
where
\begin{equation}
  x(t)\equiv -\frac{e_2t^{2/3}}{m_3^{2/3}}.
\end{equation}
The condition \reff{eq:reg0} then implies
\begin{equation}
  \label{eq:regc}
  b_1=0,\quad e_3=\frac{e_2m_4}{2m_3}=3H_0e_2.
\end{equation}
In the last equality we have used Eq.\reff{eq:m4rl}. (If we had
employed the FLRW gauge, which implies $m_4=0$, we would have had
$e_3=0$ and $b_1=0$, which is consistent with the condition derived in
\cite{VFW}.)

From Eq.\reff{eq:rho1st} we can immediately find the energy density at
the central observer. Using Eq.\reff{eq:Srl} we have
\begin{equation}
  8\pi\rho(t_0,0)=m_3.
\end{equation}
Therefore it may be natural to write
\begin{equation}
  \label{eq:defOmegas}
  \frac{m_3}{3H_0^2}=\Omega_{M,0},\quad
  \frac{e_2}{H_0^2}=\Omega_{k,0},
\end{equation}
where $\Omega_{M,0}$ and $\Omega_{k,0}$ are density parameters for the
central observer. Then we can interpret Eq.\reff{eq:e2m3} as the usual
relation for the dust FLRW model:
\begin{equation}
 \label{eq:Mkunity}
  \Omega_{M,0}+\Omega_{k,0}=1.
\end{equation}

The final task before presenting our main result is to
obtain $r(z)$ up to third order. Note that Eq.\reff{eq:drdz} is
integrable and we find
\begin{equation}
  1+z=\exp\int_0^r \frac{\dot R'(t_0-r',r')}{\sqrt{1+2E(r')}}dr'.
\end{equation}
From this equation, we have
\begin{equation}
  \label{eq:rzpan}
  \begin{split}
  r(z)= &\frac{z}{H_0}-\frac1{4H_0}\paren{5-\frac{e_2}{H_0^2}}z^2
  +\frac{1}{48 H_0}\bigg[
    109-\frac{35e_2^2+3e_4}{H_0^2e_2}+\frac{4e_2^2+e_4}{H_0^4} \\
    &
    -\frac{1}{1-H_0t_0}\paren{
      13+\frac{12b_2}{H_0}-\frac{11e_2^2+e_4}{H_0^2e_2}
      -\frac{12b_2e_2}{H_0^3}-\frac{2e_2^2-e_4}{H_0^4}
    }
  \bigg]z^3+O(z^4).
  \end{split}
\end{equation}
In particular, for $e_2=0$ we have
\begin{equation}
  \label{eq:rzpan0}
  r(z)=\frac{z}{H_0}-\frac{5}{4H_0}z^2
  +\rcp{6H_0}\paren{\frac{35}{4}-\frac{9b_2}{2H_0}-\frac{e_4}{10H_0^4}}z^3
  +O(z^4).
\end{equation}
The regularity conditions \reff{eq:regc} have been imposed in these
equations, as well as the gauge conditions \reff{eq:m4rl} and
\reff{eq:m5rl}.

We are now ready to compute the expansion of $R$ in the light cone
gauge in powers of $r$ with the regularity at $r=0$, and convert it to
in powers of $z$ using the formula \reff{eq:rzpan} or
\reff{eq:rzpan0}. As mentioned, the expansion of $R=R(t_0-r,r)$ is
straightforward if we write $R$ using the function $\S(x)$ as in
Eq.\reff{eq:RS}. Moreover, the use of Eqs.\reff{eq:Srl} and
\reff{eq:Sdrl}, together with the ODE \reff{eq:SODE2}, enables us to
eliminate the function $\S(x)$ and its derivatives
to simplify the equations.

The result of the expansion of $R$ is in such a simple form that does
not depend on the higher differential coefficients $e_4$ and $b_2$:
\begin{equation}
  \label{eq:R3rd}
  R(t_0-r,r)= r-\frac{H_0}2 r^2 
  - \frac{H_0^2}3 \paren{1-\frac{e_2}{H_0^2}}r^3 + O(r^4).
\end{equation}
Substituting Eq.\reff{eq:rzpan} and taking into account
Eq.\reff{eq:DLR}, we have the final form of regular $D_L(z)$, which is
\begin{equation}
  \label{eq:DLzreg}
  \begin{split}
  D_L(z) =&
  \frac{z}{H_0}+\frac{1}{4H_0}\paren{1+\frac{e_2}{H_0^2}}z^2 \\
  & -\frac{1}{48H_0}\bigg[
  \frac{1-e_2/H_0^2}{1-H_0t_0}
  \paren{13+\frac{2e_2}{H_0^2}+\frac{12b_2}{H_0}}
    -33 + \frac{e_2}{H_0^2}\paren{7-\frac{4e_2}{H_0^2}}
    \\
    & \hspace{5.2em}
    +  \frac1{1-H_0t_0}\paren{H_0t_0+\frac{2-3H_0t_0}{e_2/H_0^2}}
     \frac{e_4}{H_0^4}
  \bigg] z^3+O(z^4).
  \end{split}
\end{equation}
(This looks singular when $1-H_0t_0=0$, but the use of
Eqs.\reff{eq:H0} and \reff{eq:S'ineq} immediately shows $1-H_0t_0>0$.)
In particular, for $e_2=0$:
\begin{equation}
  \label{eq:DLzreg0}
  D_L(z) = \frac{z}{H_0}+\frac{z^2}{4H_0}
  -\frac{1}{8H_0}\paren{1+\frac{6b_2}{H_0}+\frac{2e_4}{15H_0^4}}z^3
  +O(z^4).
\end{equation}

It may be of some interest to have $D_L(z)$ without the regularity
conditions, which is presented in Appendix \ref{app1}.

With the above result we can easily determine an LTB model so that
its $D_L(z)$ coincides with a given luminosity distance function
$D_L^\text{(in)}(z)$ at least up to third order. The input function
$D_L^\text{(in)}(z)$ may be expanded as
\begin{equation}
  D_L^\text{(in)}(z)=I_1 z+\frac{I_2}{2}
  z^2+\frac{I_3}{3!} z^3+ \cdots.
\end{equation}
A comparison with Eq.\reff{eq:DLzreg} immediately gives
\begin{align}
  \label{eq:order1}
  H_0 &= \frac1{I_1}, \\
  \label{eq:order2}
  \frac{e_2}{H_0^2} &= 2\frac{I_2}{I_1}-1,
\end{align}
and
\begin{equation}
  \label{eq:order3}
  \begin{split}
     & \frac1{1-H_0t_0}
    \paren{H_0t_0+\frac{2-3H_0t_0}{e_2/H_0^2}} \frac{e_4}{H_0^4} 
    + \frac{1-e_2/H_0^2}{1-H_0t_0} \frac{12 b_2}{H_0}
  = \\
  & -8\frac{I_3}{I_1} +33 -\frac{e_2}{H_0^2}\paren{7-\frac{4e_2}{H_0^2}}
  - \frac{1-e_2/H_0^2}{1-H_0t_0}\paren{13+\frac{2e_2}{H_0^2}}
  \end{split}
\end{equation}
Equation \reff{eq:order1} determines $H_0$, which constrains the
parameters $e_2$, $m_3$, and $t_0$ through Eq.\reff{eq:H0}. Equation
\reff{eq:order2} determines $e_2$, which in turn determines $m_3$
through Eq.\reff{eq:e2m3}.  Then, the present time $t_0$ for the
central observer is implicitly determined from the condition
\reff{eq:Srl} with $e_2$ and $m_3$. This way we can determine the
parameters $e_2$, $m_3$, and $t_0$ for given $I_1$ and $I_2$. We can
see that the right hand side of Eq.\reff{eq:order3} is now a function
of $I_1$, $I_2$ and $I_3$. This equation represents a constraint for
$e_4$ and $b_2$, only one of which is determined from the other
through this equation. This is a result of the fact that the LTB
solution is highly degenerate in that it can represent inequivalent
multiple models that give rise to the same $D_L(z)$. If we determine
one of $e_4$ and $b_2$ according to a separate consideration, we can
determine the other. All the other parameters are determined from
Eqs.\reff{eq:m4rl}, \reff{eq:m5rl} and \reff{eq:regc}. We remark that
this procedure is solvable only for $I_2/I_1<1$ due to the inequality
\reff{eq:e2c}. (See the next section for its significance.)


\section{Comparison with the FLRW limit} 
\label{sec:4}

To understand the significance of the expansion \reff{eq:DLzreg}
itself, it is profitable to consider how much different it is from the
FLRW limit. As mentioned, one of the choices that realize the FLRW
dust solution in the LTB solution is to take
\begin{equation}
  E(r)=-\frac{k}{2} r^2=\frac{E''(0)}{2}r^2, \quad
  M(r)=M_0 r^3=\frac{M'''(0)}{3!}r^3, \quad
  t_B(r)=0.
\end{equation}
($t_B(r)$ can be any constant, for which we take zero.) However, this
coordinate choice is different from the one we have been employing. A
general gauge-independent condition to give the FLRW solution is
obtained by eliminating the explicit $r$-dependences from the above
expressions:
\begin{equation}
  \label{eq:FLRWc}
  E(r)=\frac{E''(0)}{2}\paren{\frac6{M'''(0)}M(r)}^{2/3}, \quad
  t_B(r)=0.
\end{equation}
(The constant factors have been determined so as to be valid for any
$M(r)=O(r^3)$.) The functions that satisfy both these equations and
gauge condition \reff{eq:Elgg} provide the functions that realize the
FLRW solution in the light cone gauge. Substituting the expansions
\reff{eq:fexp} into the above equations we have
\begin{equation}
  e_3=\frac{e_2m_4}{2m_3}, \quad
  e_4=\frac{e_2}{60m_3^2}\paren{-5m_4^2+24m_3m_5}.
\end{equation}
We must solve the four equations, the above two with
Eqs.\reff{eq:m4rl} and \reff{eq:m5rl}, for the four variables $e_3$,
$e_4$, $m_4$, and $m_5$. We have
\begin{equation}
  \begin{split}
    e_3 &= 3H_0 e_2,\quad e_4=(13H_0^2+2e_2)e_2, \\
    m_4 &= 6H_0 m_3, \quad m_5=5(8H_0^2+e_2)m_3,
  \end{split}
\end{equation}
which, together with $b_i=0$, give the FLRW limit for the differential
coefficients. Among the above four equations, only the equation for
$e_4$ is essential; the equation for $m_4$ is the same as the general
gauge condition \reff{eq:m4rl}, the one for $m_5$ is equivalent to the
general gauge condition \reff{eq:m5rl} with the equation for $e_4$
imposed, and the one for $e_3$ is the same as the regularity condition
\reff{eq:regc}. (The regularity at the center is therefore
automatically guaranteed for the FLRW limit as it should be.)

These limit conditions motivate us to put
\begin{equation}
  e_4=(13H_0^2+2e_2)e_2 + H_0^4\varepsilon_4,
\end{equation}
where $\varepsilon_4$ is a dimensionless parameter which becomes zero
in the FLRW limit. Substituting this into Eq.\reff{eq:DLzreg}, we have
\begin{equation}
  \label{eq:DLzreg2}
  \begin{split}
  D_L(z) = & \, D_L^{\text{(hom)}}(z) \\
  & -\frac{1}{48H_0(1-H_0t_0)}\bigg[
     \paren{H_0t_0+\frac{2-3H_0t_0}{e_2/H_0^2}} \varepsilon_4
    + \paren{1-\frac{e_2}{H_0^2}}\frac{12b_2}{H_0}
  \bigg] z^3 \\
  & +O(z^4),
  \end{split}
\end{equation}
where the $D_L(z)$ for the FLRW limit is
\begin{equation}
  \label{eq:DLzFLRW}
  D_L^{\text{(hom)}}(z) =
  \frac{z}{H_0}
  +\frac1{4H_0}\paren{1+\frac{e_2}{H_0^2}}z^2
  -\frac1{8H_0}\paren{1-\frac{e_2^2}{H_0^4}}z^3
  +O(z^4).
\end{equation}

A striking feature of our result is that it shows that \textit{the
  luminosity distance function $D_L(z)$ for an LTB solution which is
  regular at the center exactly coincides with that of an FLRW dust
  solution, up to second order.}

In particular, if we, as in \cite{BMR05}, define the deceleration
parameter $q_0$ in comparison with the $D_L(z)$ for a general FLRW
model:
\begin{equation}
  \label{eq:defq0}
  D_L(z)=\rcp{H_0}\paren{z+\frac{1-q_0}{2}z^2+O(z^3)},
\end{equation}
we have $q_0>0$, as in the FLRW dust solution. In fact, comparing with
Eq.\reff{eq:DLzreg} we have
\begin{equation}
  \label{eq:q0}
  q_0=\rcp2\paren{1-\frac{e_2}{H_0^2}}>0,
\end{equation}
because of Eq.\reff{eq:e2c}. This reconfirms the claim in
\cite{VFW,Fl,HS}.

\section{Conclusions}
\label{sec:conc}

We have first given a new way of expressing the LTB solution, which is
explicit (i.e., not parametric) and requires no separate
considerations depending on the sign of the energy function
$E(r)$. This has been done using a ``special'' function $\S(x)$, which
can be defined as the unique `transversal' solution of the first order
ODE \reff{eq:SODE1}. Using this monotonic function, we can write the
areal radius function $R(t,r)$ for the LTB metric in the concise form
\reff{eq:RS}. To simplify expressions involving higher derivatives of
$\S(x)$ it is most useful to use the second order ODE \reff{eq:SODE2}.

Taking advantage of this concise expression, we have computed the
luminosity distance function $D_L(z)$ for the LTB solution up to third
order of $z$. We have found that if we impose the regularity condition
at the center, the function degenerates into an FLRW dust case up to
second order, and differences only appear from the third order.

The second order coincidence with the FLRW dust solution tells us that
we cannot choose a set of LTB functions $\{E(r),t_B(r),M(r)\}$ so that
the resulting $D_L(z)$ fits an FLRW model which shows an accelerating
expansion like the $\Lambda$CDM model, as long as the regularity
condition is imposed. This is however of course not to say that we
cannot find an LTB model that explains the observations, since for
this it is not necessary to have an exact fitting of $D_L(z)$ near the
center with an accelerating FLRW model.

Perhaps, the simplest way to guarantee the regularity and still have a
good fit of $D_L(z)$ with the observations is to choose the model so
that it \textit{exactly} coincides with an FLRW dust solution (of
perhaps negative curvature) for $z$ smaller than a certain small (but
finite) value $z_1$, and use the full flexibility of the LTB solution
for $z>z_1$ to fit the observations. This approach however has the
drawback that the $D_L(z)$ chosen this way inevitably becomes a
non-analytic function, since if it was analytic it would be that of
the FLRW solution.  To be sure, it is possible to approximate such a
non-analytic function with an appropriate analytic function. For
example, one might use $\tanh x$ to approximate a step function, but
$\tanh x$ never becomes constant for large $x$, as opposed to the step
function. Like this, an analytic $D_L(z)$ chosen to approximate a
$D_L(z)$ that is endowed with the above property deviates from that of
FLRW near the center.

To summarize, if one wants $D_L(z)$ to be analytic (or of any
arbitrary form), the inverse problem at the center is nontrivial. Our
result \reff{eq:DLzreg} gives, as we have seen, a solution to this
problem.  A further study based on the present work is under progress.

\section*{Acknowledgments}

We thank Satoshi Gonda for stimulating discussions.

\appendix

\section{$D_L(z)$ without regularity condition}
\label{app1}

In this Appendix, we present the luminosity distance function for the
LTB solution that is not necessarily regular at the center. To
distinguish from the regular one, we write $D_L^\textrm{(all)}(z)$ to
denote this general function. Let $D_L(z)$ be the regular part of
$D_L^\textrm{(all)}(z)$ as in Eq.\reff{eq:DLzreg} or
\reff{eq:DLzreg2}, and $d_L(z)$ be the difference from
the regular part, i.e.,
\begin{equation}
  D_L^\textrm{(all)}(z)=D_L(z)+d_L(z).
\end{equation}
The difference part $d_L(z)$ should vanish when the
conditions \reff{eq:regc} are satisfied. Motivated by the condition for
$e_3$, we rewrite $e_3$ using the dimensionless parameter
$\varepsilon_3$ defined by
\begin{equation}
  e_3=3H_0e_2+H_0^3\varepsilon_3.
\end{equation}
The parameter $\varepsilon_3$ vanishes when $e_3$ satisfies the
regularity condition.

Then, we have
\begin{equation}
  d_L(z) =
  \frac{-A_2}{12H_0(1-H_0t_0)}  z^2
  +\frac{{A}_3}{288 H_0(1-H_0t_0)^3}  z^3
  +O(z^4),
\end{equation}
where
\begin{equation}
  {A}_2\equiv
    \paren{H_0t_0+\frac{2-3H_0t_0}{e_2/H_0^2}} \varepsilon_3
    +6\paren{1 - \frac{e_2}{H_0^2}}b_1 ,
\end{equation}
and
\begin{equation}
  \begin{split}
    {A}_3 \equiv & \,
      36(4-5 H_0t_0) \paren{1-\ev}^2  b_1^2 \\
      & + 12 \bigg[ 2H_0t_0(8 -7 H_0t_0)-4
      +\ev\paren{\paren{\patLX}^2-H_0t_0(4-5H_0t_0)} \bigg]
      \, b_1 \varepsilon_3 \\
      & + \bigg[ H_0^2t_0^2(4-5H_0t_0)
      -\frac25 \paren{\patLX}(5-H_0t_0(21-20 H_0t_0)) \\
      & 
      +\paren{\patLX+\frac25}\frac{10-H_0t_0(22-13H_0t_0)}{\eh} \bigg]
      \, \varepsilon_3^2
       \\
      &+ 4(\patT)\bigg\{
       6 \paren{1-\ev}
       \paren{9-11H_0t_0-2(3-4 H_0t_0)\ev} \, b_1 \\
      & + \bigg[
      \patLX(15-17 H_0t_0) 
      +33H_0t_0(1-H_0t_0) \\
      &  -4 -2H_0t_0(3-4H_0t_0)\ev
      \bigg] \, \varepsilon_3
      \bigg\}.
  \end{split}
\end{equation}
To take the limit $e_2\goes0$ of this expression, one may need both of
Eqs.\reff{eq:e20limiteqn} and \reff{eq:e20limiteqn2}.  As a result,
for $e_2=0$ we have
\begin{equation}
  \label{eq:DLzd0}
  d_L(z) =
  -\frac{B_2}{H_0}  z^2
  +\frac{B_3}{H_0}  z^3
  +O(z^4),
\end{equation}
where
\begin{equation}
  B_2\equiv \frac{\varepsilon_3}{15}+\frac{3b_1}{2},
\end{equation}
and
\begin{equation}
  B_3\equiv
    \frac{7\,\varepsilon_3}{30} 
    + \frac{13\,{\varepsilon_3}^2}{1575} 
    + \frac{\varepsilon_3\,b_1}{2} 
    + \frac{5\,b_1}{4} 
    + \frac{9\,{b_1}^2}{4}.
\end{equation}

The deceleration parameter $q_0$ for the central observer is therefore
\begin{equation}
  q_0=\rcp2\paren{1-\ev}+\frac1{6(\patT)}
  \bra{\paren{H_0t_0+\frac{2-3H_0t_0}{\eh}}\varepsilon_3
    +6\paren{1-\ev}b_1}.
\end{equation}
In particular, for the $e_2=0$ case,
\begin{equation}
  q_0=\rcp2+\frac2{15}\varepsilon_3+3b_1,
\end{equation}
which is apparently positive unless at least one of $\varepsilon_3$ or
$b_1$ is nonzero. This is also the case for any $e_2$ including
$e_2\neq0$, as remarked in Section \ref{sec:4}. Nonzero
$\varepsilon_3$ or $b_1$ however leads to a weak singularity at $z=0$.

We comment that our $D_L^\textrm{(all)}(z)$ is a generalization of the
result of C\'{e}l\'{e}rier \cite{Ce} which corresponds to the case
$e_i=0$. To confirm the consistency, however, there are two remarks to
make; (i) Ref.\cite{Ce} employs the FLRW gauge, which is different
from ours. To compare, an appropriate transformation is needed between
the differential coefficients of the three LTB functions. (ii) Eq.(45)
of \cite{Ce} contains two wrong numerical factors; $7$ and $10$ should
read, respectively, $6$ and $9$. We have confirmed that our result is
consistent with \cite{Ce} after taking these two points into account.


\begin{thebibliography}{AA}

\bibitem{Riess:1998cb}
  A.~G.~Riess {\it et al.}  [Supernova Search Team Collaboration],
  Astron.\ J.\  {\bf 116}, 1009 (1998)
  [arXiv:astro-ph/9805201].

\bibitem{Perlmutter:1998np}
  S.~Perlmutter {\it et al.} [Supernova Cosmology Project Collaboration],
  Astrophys.\ J.\  {\bf 517}, 565 (1999)
  [arXiv:astro-ph/9812133].

\bibitem{Riess:2004nr}
  A.~G.~Riess {\it et al.}  [Supernova Search Team Collaboration],
  Astrophys.\ J.\  {\bf 607}, 665 (2004)
  [arXiv:astro-ph/0402512].

\bibitem{Spergel:2003cb}
  D.~N.~Spergel {\it et al.} [WMAP Collaboration],
  Astrophys.\ J.\ Suppl.\  {\bf 148}, 175 (2003)
  [arXiv:astro-ph/0302209].

\bibitem{Tegmark:2003ud}
  M.~Tegmark {\it et al.}  [SDSS Collaboration],
  Phys.\ Rev.\  D {\bf 69}, 103501 (2004)
  [arXiv:astro-ph/0310723].

\bibitem{L}
  G.~Lema\^itre,
  Ann.\ Soc.\ Sci.\ Bruxelles A {\bf 53}, 51 (1933);
  reprinted in  Gen.\ Rel.\ Grav.\ {\bf 29}, 641 (1997).

\bibitem{T}
  R.~C.~Tolman,
  Proc.\ Nat.\ Acad.\ Sci.\ {\bf 20}, 169 (1934);
  reprinted in  Gen.\ Rel.\ Grav.\ {\bf 29}, 935 (1997).

\bibitem{B}
  H.~Bondi,
  Mon.\ Not.\ R.\ Astron.\ Soc.\ {\bf 107}, 410 (1947).

\bibitem{Ce}
  M.~-N.~C\'{e}l\'{e}rier,
  Astron.\ Astrophys.\  {\bf 353}, 63 (2000)
  [arXiv:astro-ph/9907206].

\bibitem{Iguchi:2001sq}
  H.~Iguchi, T.~Nakamura and K.~i.~Nakao,
  Prog.\ Theor.\ Phys.\  {\bf 108}, 809 (2002)
  [arXiv:astro-ph/0112419].

\bibitem{Alnes:2005rw}
  H.~Alnes, M.~Amarzguioui and O.~Gron,
  Phys.\ Rev.\  D {\bf 73}, 083519 (2006)
  [arXiv:astro-ph/0512006].

\bibitem{Bolejko:2005fp}
  K.~Bolejko,
  arXiv:astro-ph/0512103.

\bibitem{Garfinkle:2006sb}
  D.~Garfinkle,
  Class.\ Quant.\ Grav.\  {\bf 23}, 4811 (2006)
  [arXiv:gr-qc/0605088].

\bibitem{Biswas:2006ub}
  T.~Biswas, R.~Mansouri and A.~Notari,
  arXiv:astro-ph/0606703.
  
\bibitem{Enqvist:2006cg}
  K.~Enqvist and T.~Mattsson,
  JCAP {\bf 0702}, 019 (2007)
  [arXiv:astro-ph/0609120].

\bibitem{El98}
  N.~Mustapha, C.~Hellaby and G.~F.~R.~Ellis,
  Mon.\ Not.\ Roy.\ Astron.\ Soc.\  {\bf 292}, 817 (1997)
  [arXiv:gr-qc/9808079].

\bibitem{VFW}
  R.~A.~Vanderveld, \'{E}.~\'{E}.~Flanagan, and I.~Wasserman,
  Phys.\ Rev.\ D {\bf 74}, 023506 (2006)
  [arXiv:astro-ph/0602476].

\bibitem{Nambu:1999aw}
  Y.~Nambu and Y.~Y.~Yamaguchi,
  Phys.\ Rev.\  D {\bf 60}, 104011 (1999)
  [arXiv:gr-qc/9904053].

\bibitem{Ellis:1971pro}
G.~F.~R.~Ellis, in ``General Relativity and Cosmology,''
Proc. Int. School of Physics ``Enrico Fermi'', Course XLVII,
ed. R.~K.~Sachs, Academic Press (1971) 104.

\bibitem{PM84} 
  M.~H.~Partovi and B.~Mashhoon,
  Astrophys.\ J.\  {\bf 276}, 4 (1984).

\bibitem{El97}
  N.~Mustapha, B.~A.~Bassett, C.~Hellaby and G.~F.~R.~Ellis,
  Class.\ Quant.\ Grav.\  {\bf 15}, 2363 (1998)
  [arXiv:gr-qc/9708043].

\bibitem{Mustapha:2000bf}
  N.~Mustapha and C.~Hellaby,
  Gen.\ Rel.\ Grav.\  {\bf 33}, 455 (2001)
  [arXiv:astro-ph/0006083].

\bibitem{BMR05}
  E.~Barausse, S.~Matarrese and A.~Riotto,
  Phys.\ Rev.\  D {\bf 71}, 063537 (2005)
  [arXiv:astro-ph/0501152].

\bibitem{Fl}
  \'{E}.~\'{E}.~Flanagan,
  Phys.\ Rev.\  D {\bf 71}, 103521 (2005)
  [arXiv:hep-th/0503202].

\bibitem{HS}
  C.~M.~Hirata and U.~Seljak,
  Phys.\ Rev.\  D {\bf 72}, 083501 (2005)
  [arXiv:astro-ph/0503582].

  
\end{thebibliography}
\end{document}